\documentclass[14pt,letterpaper,times]{article}
\usepackage[tcidvi]{graphicx}
\usepackage[english]{babel}
\usepackage{amsmath,amsfonts,graphics,color,colortbl,fontsmpl,amssymb,amsfonts,epsfig,rotating,float,amscd,multicol,layout}

\begin{document}
\title{Negativity and Concurrence as complete entanglement measures for two arbitrary qudits}
\author{Suranjana Rai$^\dagger$\quad and Jagdish R. Luthra\footnote{e-mail:
jluthra@uniandes.edu.co}\\
\small{($\dagger$) Raitech, Tuscaloosa, AL 35405}\\
\small{($*$) Departamento de F\'isica, Universidad de los Andes,
A.A. 4976 Bogot\'a, Colombia}}
\date{August 5, 2005}
%
\maketitle
\begin{abstract}
Two measures of entanglement, negativity and concurrence are studied
for two arbitrary qudits. We obtain negativity as an expectation
value of an operator. The differences of the squares of negativity
and concurrence are invariants of multilevel entanglement. Explicit
results for qutrits and quadrits are obtained.\\
\end{abstract}

For two qubits the defining measure of entanglement is concurrence.
It is a good measure of entanglement in  every sense. It is
computable, it does not change under local operations and classical
communications (LOCC) and it gives rise to the entanglement of
formation for pure and mixed states \cite{wootters, rungta,
uhlmann}. So far the defining measure or set of measures has not
been found for arbitrary qudits. In this letter we show that the
negativity, $N$, and the concurrence, $C$, along with the difference
of their squares are good candidates for complete description of
entanglement for bipartite qudits.\\


For qubits, Wootters \cite{wootters} defines concurrence using the
Pauli spin matrix $\sigma_y$ as a spin flip operator. For qubits,
$\sigma_{y}$ transforms maximally entangled states (Bell states)
to themselves, so that the concurrence for these states is one.
This is possible because of the special symmetry of $2\otimes2$
systems. In general, for bipartite states of arbitrary dimension,
instead of the Bell states, we use states in the Schmidt form. All
of the Bell states are not in the Schmidt form as the Schmidt
coefficients have to be real and non-negative. A general two qubit
pure state in the Schmidt form is
\begin{eqnarray}
|\psi\rangle &=& \sum_{i=1}^2 k_i|ii\rangle
\end{eqnarray}
where $k_1,k_2$ are the two Schmidt coefficients with the
normalization condition
\begin{eqnarray}
k_1^2 +k_2^2 &=& 1
\end{eqnarray}

The concurrence as defined by Wootters \cite{wootters} is
\begin{eqnarray}
C(|\psi\rangle) &=&
\langle\psi|\sigma_{y}\otimes\sigma_{y}|\psi^{*}\rangle
\end{eqnarray}
where $\psi^*$ denotes the complex conjugation and $\sigma_y$ is one
of Pauli spin operators. For the state in Eq.(1), the concurrence is
\begin{eqnarray}
C &=& 2k_1k_2
\end{eqnarray}
The above concurrence can also be written as \cite{rungta}
\begin{eqnarray}
C = \sqrt{2(1-Tr\rho_A^2)}=2k_1k_2
\end{eqnarray}
where $\rho_A$ is the reduced density matrix.\\

Another important measure of entanglement is the negativity
\cite{horodecki, jens, vidal,myungshik, rai}. Negativity is an
entanglement monotone so it does not change under LOCC. The
negativity \cite{vidal} of a bipartite system described by the
density matrix $\rho$, is given by $N(\rho)$ as
\begin{eqnarray}
N(\rho) &=& \frac{||\rho^{T_A}||_1 - 1}{2}\
\end{eqnarray}
where $\rho^{T_A}$ is the partial transpose with respect to system
A, and $||...||_1$ denotes the trace norm. The negativity is a
quantitative measure of the partial positive transpose (PPT), the
Peres criteria \cite{peres}. It measures how negative the
eigenvalues of the density matrix are after the partial transpose is
taken. It is the absolute value of the sum of the negative
eigenvalues of the partially transposed density matrix. According to
the Peres criterion of separability, density matrices with NPT are
entangled \cite{peres}. The Peres criterion is necessary and
sufficient for qubit-qubit and qubit-qutrit systems. For systems of
higher dimensions, it is necessary but not sufficient, since, there
are states with PPT which are entangled. These states are said to
have bound entanglement since entanglement cannot be distilled from
these states. The negativity can be generalized to higher dimensions
\cite{lee} as
\begin{eqnarray}
N(\rho) &=& \frac{||\rho^{T_A}||_1 -1}{d-1}
\end{eqnarray}
where $d$ is the smaller of the dimensions of the bipartite system.
For two qubits, the negativity is
\begin{eqnarray}
N&=& 2k_1k_2
\end{eqnarray}
Now we show that the negativity can be obtained from the action of
an operator $X$ defined as
\begin{eqnarray}
X = X_1\otimes X_2 = \sigma_x^1\otimes\sigma_x^2
\end{eqnarray}
where $X_1=\sigma_x^1$ is the Pauli operator that acts on the first
qubit and $X_2=\sigma_x^2$ acts on the second qubit.
\begin{eqnarray}
X|11\rangle =|22\rangle ,    X|22\rangle = |11\rangle
\end{eqnarray}
We define the negativity as the expectation value of the operator
$X$ in the state $|\psi\rangle$.
\begin{eqnarray}
\langle X \rangle =2k_1k_2=N(|\psi\rangle)
\end{eqnarray}
For two qubits in the Schmidt form, we see from Eqs.(4), (8) and
(11), that the negativity and the concurrence turn out to be
identical, i.e.,
\begin{eqnarray}
N &=& C
\end{eqnarray}
For mixed states, concurrence can be extended by the convex roof
\cite{wootters, chen},
\begin{eqnarray}
C(\rho)\equiv min\sum_ip_i C(|\psi_i\rangle)
\end{eqnarray}
where the minimum is taken over the ensemble of all possible
decompositions of the density matrix $\rho$. The negativity can also
be extended by a similar procedure \cite{lee}.
\begin{eqnarray}
N_m(\rho)\equiv min\sum_ip_i N(|\psi_i\rangle)
\end{eqnarray}
Lee et. al.,\cite{lee} have also shown that for a mixed two qubit
case, the negativity is equal to the concurrence. They state that
the convex roof extended negativity is better at detecting
separability than the original negativity.\\

Now, we would like to generalize to arbitrary bipartite qudit
states. We consider the general case of $m\otimes n$ and $d =
min(m,n)$. The general bipartite pure state in the Schmidt form can
be written as
\begin{eqnarray}
|\psi\rangle &=& \sum_{i=1}^d k_i|ii\rangle
\end{eqnarray}
with the usual normalization condition
\begin{eqnarray}
 \sum_{i} k_i^2 = 1
\end{eqnarray}
The generalized Pauli spin operator $X$ \cite{jay} acts as
\begin{eqnarray}
X|i,i\rangle =|i+1,i+1\rangle, mod(d)
\end{eqnarray}
The $cyclic$ operator $X$ transforms the state $|\psi\rangle$ into a
shifted state by the ladder action. $X$, is unitary but not
hermitian. Now for pure states in the Schmidt form, using $X$ is
sufficient as the Schmidt coefficients are real. In general, if we
do not use the Schmidt form, the hermitian operators of the form $(X
\pm X^\dag)$ have to be used \cite{cesar}. We propose the use of the
generalized operator $X$ to connect the different levels of the
states. It is easy to see that negativity can still be obtained as
the expectation value of the $X$ operator in the state
$|\psi\rangle$.
\begin{eqnarray}
N(|\psi\rangle)=\langle X \rangle =\sum_{i<j} k_i k_j
\end{eqnarray}
On the other hand, the concurrence for the general state Eq.(15) is
\begin{eqnarray}
C(|\psi\rangle)=\sqrt{2(1-Tr\rho_A^2)} =\sqrt{4\sum_{i<j} k_i^2
k_j^2}
\end{eqnarray}
Chen et. al., \cite{chen} have shown that
\begin{eqnarray}
4\sum_{i<j} k_i^2 k_j^2\geq \frac{2}{d(d-1)}(\sum_{i<j} k_i k_j)^2
\end{eqnarray}
We recognize the terms on the left and the right in terms of the
concurrence and negativity respectively. It is important to take
care of scaling in Eq.(20). For two qubits the maximum value of $N$
and $C$ are equal to $1$. However, for systems of higher dimensions,
the maximum value of concurrence and negativity on this scale depend
on the dimensions of the system. For $d=3$, the maximum value of $C$
is $\frac{2}{\sqrt{3}}$. For $d = 4$, the maximum value is
$\sqrt{\frac{3}{2}}$. For general $d$, the maximum value is
$\sqrt{2(1-\frac{1}{d})}$. As $d$ becomes very large, this converges
to a value of $\sqrt{2}$. However, on this scale for arbitrary $d$,
when we consider the subspace of states with only two nonzero
Schmidt coefficients, these states have a maximum value of
concurrence as one. Of course, the completely separable states have
concurrence zero. We also rescale the negativity to have a maximum
value of one in any dimension. Therefore the rescaled negativity is
\begin{eqnarray}
N(|\psi\rangle)= \frac{2}{(d-1)}\sum_{i<j} k_i k_j
\end{eqnarray}
This negativity still has the value one for two qubits. Eq.(20) is
now restated in terms of the concurrence and the negativity as
\begin{eqnarray}
C^2 \geq \frac{d-1}{2d}N^2
\end{eqnarray}

It is interesting to see specific relations between N and C in
higher dimensions. We have explored this connection for two qutrits
and given an analytic expression relating the two quantities
\cite{rai}. We also showed that for two qutrits concurrence is lower
bounded by negativity. For two qutrits the relation between the
concurrence and the rescaled negativity is
\begin{eqnarray}
N^2 &=& \frac{C^2}{4} \pm 2(k_1k_2k_3)\sqrt{1+2N}
\end{eqnarray}
In the previous paper \cite{rai}, we had given a relation between
$N$ and $C$ for qutrits where $C$ was also scaled to one. Both
concurrence and negativity give entanglement as a sum of
entanglement between pairs of levels. Concurrence is the root mean
square of pair-wise products of the Schmidt coefficients while
negativity is the sum of pair-wise products of the coefficients. The
difference contains a part which occurs with the product of the
three Schmidt coefficients, $k_1k_2k_3$. This product vanishes when
three way entanglement is absent,i.e., when one of the Schmidt
coefficient is zero, in which case $N = C/2$. For maximum
entanglement, $C=1$ and $N=1/2$. It is significant to see that the
difference between $C$ and $N$ is in terms of the three-level
entanglement of the two qutrits. The product $k_1k_2k_3$ is related
to the determinant of the coefficient matrix, which is an invariant
under local unitaries.\\

On similar lines, we would like to extend this procedure to
quadrits. The difference of the squares of negativity and
concurrence generates interesting quantities that could be useful in
the description of various types of entanglements possible. Now we
relate the square of the negativity of two quadrits to the
concurrence. The negativity and concurrence for two quadrits are
\begin{eqnarray}
N = \frac{2}{3}(k_1k_2 + k_1k_3+k_1k_4+k_2k_3+k_2k_4+k_3k_4)\\\
C=4(k_1^2k_2^2 +
k_1^2k_3^2+k_1^2k_4^2+k_2^2k_3^2+k_2^2k_4^2+k_3^2k_4^2)
\end{eqnarray}
We invoke some results from the theory of entanglement and
invariants \cite{grassl}to understand the various terms that arise
while relating $N$ and $C$. For quadrits there are four symmetric
polynomials that are invariant under unitary transformations:
\begin{eqnarray}
s_1 = k_1 +k_2+k_3+k_4 \\\
s_2 =k_1k_2 + k_1k_3+k_1k_4+k_2k_3+k_2k_4+k_3k_4\\
s_3 = k_1k_2k_3+k_2k_3k_4+k_3k_4k_1+k_4k_1k_2\\
s_4 =k_1k_2k_3k_4
\end{eqnarray}
The invariants in the above equations are given in terms of Schmidt
coefficients. Negativity has a very simple structure in terms of
pairs of Schmidt coefficients and is the invariant $s_2$.
\begin{eqnarray}
N = \frac{2}{3}s_2
\end{eqnarray}
We see that negativity contains entirely pair-wise entanglement. The
concurrence on the hand, involves other invariants which are related
to three-level and four-level entanglements through invariants $s_1,
s_3 and s_4$. The square of concurrence can be written as
\begin{eqnarray}
C^2 = 4s_2^2 +2(s_4-s_1s_3)
\end{eqnarray}
The difference measures the contribution to entanglement from a
larger number of level entanglements. In fact we can obtain a
general relation connecting the concurrence to negativity as
\begin{eqnarray}
C^2 - 4 N^2 = 2(s_4 - s_1s_3)
\end{eqnarray}

We could continue to explore the squares of negativity and
concurrence in arbitrary dimensions. We conclude by making some
general observations about higher dimensional qudits. Since the
negativity and concurrence are both invariants, the difference
between them in the general case is also an invariant. We can study
invariant multilevel entanglement from these quantities. Higher
powers would provide complete information about entanglement. These
results can also be extended to mixed states.

\end{document}